\documentclass[aps,preprint,nofootinbib,superscriptaddress,showpacs]{revtex4-1}

\usepackage{amsmath}
\usepackage{amssymb}
\usepackage{mathrsfs}
\usepackage[breaklinks]{hyperref}
\usepackage{graphicx}
\usepackage{CJK}

\begin{document}
%
\begin{CJK}{UTF8}{gkai}
%
\title{$CP$ violation  induced by the interference of scalar and vector resonances in three-body decays of bottom mesons}
%
%
\author{Zhen-Hua~Zhang (张振华)}
 \email[Email:]{zhangzh@iopp.ccnu.edu.cn}
 \affiliation{School of Nuclear Science and Technology, University of South China, Hengyang, Hunan 421001, China}
\author{Xin-Heng~Guo (郭新恒)}
\email[Email: ]{xhguo@bnu.edu.cn}
\affiliation{College of Nuclear Science and Technology, Beijing Normal University, Beijing 100875, China}%
\author{Ya-Dong~Yang (杨亚东)}
\email[Email:]{yangyd@iopp.ccnu.edu.cn}\affiliation{Institute of Particle Physics, Huazhong Normal University, Wuhan 430079, China}
 \date{\today}
%
\begin{abstract}
Large $CP$ violation is an interesting phenomenon both theoretically and experimentally.
Last year, LHCb Collaboration found in some three-body decays of bottom mesons that large $CP$ violations appear in regions of the Dalitz plots that are not dominated by contributions from narrow resonances.
In this paper, we present a mechanism which can induce such kind of large $CP$ violations.
In this mechanism, large localized $CP$ asymmetries in phase space can be induced  by the interference of two intermediate resonances with different spins.
We also apply this mechanism to the decay channel $B^\pm\to K^\pm \pi^+\pi^-$.
\end{abstract}
%
%
\maketitle
\end{CJK}
%
%
\section{introduction}
%
Charge-Parity ($CP$) violation is one of the most fundamental and important properties of weak interactions.
It was first discovered in $K^0-\overline{K^0}$ systems in 1964 \cite{Christenson:1964fg}.
In Standard Model (SM), $CP$ violation is originated from the weak phase in Cabibbo-Kobayashi-Maskawa (CKM) matrix which describes the mixing of different generations of quarks \cite{Cabibbo:1963yz,Kobayashi:1973fv}.
Besides the weak phase, in order to have a $CP$ asymmetry that is large enough to detect, a large strong phase is needed. 
Usually, this large phase is provided by QCD loop corrections.

It was suggested long time ago that large $CP$ violation should be observed in $B$ meson systems \cite{Carter:1980tk,Bigi:1981qs}.
Last year, LHCb Collaboration found clear evidence for $CP$ violation in some three-body decay channels of $B$ mesons \cite{LHCb-CONF-2012-018,LHCb-CONF-2012-028,deMiranda:2013kg}.
Intriguingly, large direct $CP$ asymmetries were found in some localized phase spaces of the decay channel $B^\pm\to\pi^\pm\pi^+\pi^-$, which dose not clearly correspond to any resonance \cite{LHCb-CONF-2012-028,deMiranda:2013kg}.
The observed large localized $CP$ asymmetry lies in the region  $m^2_{\pi^+\pi^-~\text{low}}<0.4~\text{GeV}^2$ and $m^2_{\pi^+\pi^-~\text{high}}>15~\text{GeV}^2$
\footnote{For the decay channel $B^-\to \pi^-\pi^+\pi^-$, there are two identical  pions with negative charge.
When combining the momentum of each $\pi^-$ meson with that of the $\pi^+$ meson, we will have two Lorentz invariant masses squared which are usually different in values and are denoted as $m_{\pi^+\pi^-~\text{low}}^2$ and $m_{\pi^+\pi^-~\text{high}}^2$ in Ref. \cite{deMiranda:2013kg}, respectively.}%
, and takes the value:
\begin{equation}
 A_{CP}=+0.622\pm0.075\pm0.032\pm0.007,
\end{equation}
 while in the region $m^2_{\pi^+\pi^-~\text{low}}<0.4~\text{GeV}^2$ and $m^2_{\pi^+\pi^-~\text{high}}<15~\text{GeV}^2$, no large $CP$ asymmetry is observed.

In our previous paper \cite{ZhangGuoYang}, we proposed a mechanism which can generate large localized $CP$ asymmetries in phase space of three-body decay by the interference of two intermediate resonances with different spins.
With this mechanism, we showed that the large $CP$ asymmetry difference between the aforementioned two regions can be interpreted as the interference  of amplitudes which correspond to two intermediate resonances, $\rho^0(770)$ and $f_0(500)$, respectively.

In fact, similar $CP$ asymmetry behavior was also observed in $B^\pm\to K^\pm\pi^+\pi^-$ \cite{LHCb-CONF-2012-018,deMiranda:2013kg}.
When the invariant mass of the $\pi^+\pi^-$ pair is around the vicinity of $f_0(500)$, a $CP$ asymmetry larger than about 30\% was observed for smaller invariant mass of the $K^\mp\pi^\pm$ system, while a $CP$ asymmetry that is slightly smaller than 0 (about 0 to -10\%) was observed for larger invariant mass of the $K^\mp\pi^\pm$ system.
In this paper, we will first give a more general analysis of the aforementioned mechanism, and then apply it to the channel $B^\pm\to K^\pm\pi^+\pi^-$.

The remainder of this paper is organized as follows. 
In Sec. \ref{sec:GeneralConsideration}, we first present a detailed analysis of the aforementioned mechanism which can be generate large localized $CP$ asymmetries in three-body decays of bottom mesons.
In Sec. \ref{sec:Application},  we apply the mechanism to the decay channel $B^\pm\to K^\pm\pi^+\pi^-$.
In Sec. \ref{sec:Conclusion}, we present our conclusions.

%
\section{\label{sec:GeneralConsideration} General Consideration on the Interference of Two nearby resonances with different spins}
%
For a cascade decay process, $B\to X M_3$, $X\to M_1 M_2$, with all the initial and final particles being spin-0 ones, the transition amplitude is proportional to $P_{J_X}\big(g_{s_{12}}(s_{13})\big)$ \cite{Chung:1971ri}, where $P_{J_X}$ is the $(J_X+1)$-th Legendre Polynomial, $s_{ij}$ ($i,j=1,2,3$) is the invariant mass squared of $M_i$ and $M_j$, $J_X$ is the spin of $X$, and
\begin{equation}
g_{s_{12}}(s_{13})=\frac{\hat{s}_{13}-s_{13}}{\Delta_{13}},
\end{equation}
with $\hat{s}_{13}=(s_{13,\text{max}}+s_{13,\text{min}})/2$, $\Delta_{13}=(s_{13,\text{max}}-s_{13,\text{min}})/2$, and $s_{13,\text{max(min)}}$ being the maximum (minimum) value of $s_{13}$ for fixed $s_{12}$.

Inspired by the above statement,  we can expand the transition amplitude of the decay process, $B\to M_1 M_2 M_3$,  in terms of Legendre polynomials for fixed $s_{12}$:
\begin{equation}
\mathcal{M}(s_{12}, s_{13})=\sum_{l} a_{l}P_l\big(g_{s_{12}}(s_{13})\big).
\end{equation}
Note that $a_l$, $\Delta_{13}$, and $\hat{s}_{13}$ may depend on $s_{12}$, but all of them are independent of $s_{13}$.
For certain value of $s_{12}$ (denoted by $\bar{s}_{12}$) when $a_J$ is much larger than other $a_l$'s, the transition amplitude $\mathcal{M}$ will be dominated by the $(J+1)$-th Legendre Polynomial:
\begin{equation}
\mathcal{M}(\bar{s}_{12}, s_{13})\simeq a_JP_{J}\big(g_{\bar{s}_{12}}(s_{13})\big).
\end{equation}
One would observe a spin-$J$ resonance lying around $s_{12}=\bar{s}_{12}$, which is in fact responsible for the aforementioned cascade decay.

Another interesting situation arises when two different Legendre Polynomials with $l=J_1$ and $l=J_2$ are dominant for fixed $s_{12}=\bar{s}_{12}$.
The decay amplitude $\mathcal{M}$ will take the form
\begin{equation}
\mathcal{M}(\bar{s}_{12}, s_{13})\simeq a_{J_1}P_{J_1}\big(g_{\bar{s}_{12}}(s_{13})\big)+a_{J_2}P_{J_2}\big(g_{\bar{s}_{12}}(s_{13})\big).
\end{equation}
If this decay process is a weak one, $a_l$'s may take a general form
\begin{equation}
a_{l}=\left[\mathcal{T}_l+\mathcal{P}_l e^{i(\alpha_l+\phi)}\right]e^{i\delta_l},
\end{equation}
where $\phi$ is the weak phase, while $\delta_l$ and $\alpha_l$ are strong phases, $\mathcal{T}_l$ and $\mathcal{P}_l$ represent tree and penguin amplitudes, respectively.
The strong phases $\delta_l$ and $\alpha_l$ can be properly chosen so that both $\mathcal{T}_l$ and $\mathcal{P}_l$ are real.
The differential $CP$ violation parameter, which is defined as 
\begin{equation}
A_{CP}=\frac{|\mathcal{M}|^{2}-|\overline{\mathcal{M}}|^{2}}{|\mathcal{M}|^{2}+|\overline{\mathcal{M}}|^{2}},
\end{equation}
can then be expressed as $A_{CP}=D/F$, where
\begin{eqnarray}
D&=&-2\sin\phi\Big\{\Big[P_{J_1}P_{J_2}\mathcal{T}_{J_1}\mathcal{P}_{J_2}\sin(\delta_{J_2}
-\delta_{J_1}+\alpha_{J_2})
+P_{J_1}^2\mathcal{T}_{J_1}\mathcal{P}_{J_1}\sin\alpha_{J_1}\Big]+[J_1\leftrightarrow J_2]\Big\}%
,%
\\ 
F&=&\Big\{P_{J_1}^2(\mathcal{T}_{J_1}^2+\mathcal{P}_{J_1}^2)
+P_{J_1}P_{J_2}\Big[\mathcal{T}_{J_1}\mathcal{T}_{J_2}\cos(\delta_{J_1}-\delta_{J_2})
+\mathcal{P}_{J_1}\mathcal{P}_{J_2}\cos(\delta_{J_1}-\delta_{J_2}+\alpha_{J_1}-\alpha_{J_2})\Big]
\nonumber\\&&
+2\cos\phi\Big[P_{J_1}P_{J_2}\mathcal{T}_{J_1}\mathcal{P}_{J_2}\cos(\delta_{J_2}-\delta_{J_1}+\alpha_{J_2})
+P_{J_1}^2\mathcal{T}_{J_1}\mathcal{P}_{J_1}\cos\alpha_{J_1}\Big]\Big\}+\{J_1\leftrightarrow J_2\}%
,
\end{eqnarray}
with $P_{J_i}$ $(i=1,2)$ being the abbreviation for $P_{J_i}\big(g_{\bar{s}_{12}}(s_{13})\big)$.
One can see that the $CP$ asymmetry depends on $s_{13}$ through $P_{J_1}$ and $P_{J_2}$.
This is a very interesting behavior.
On the other hand, no $s_{13}$-dependence of the $CP$ asymmetry appears if only one Legendre Polynomial dominates because the common factor $P_J^2$ will be cancelled between $D$ and $F$.
When $\alpha_1$ and $\alpha_2$ equal zero, only one strong phase $\delta\equiv(\delta_{J_1}-\delta_{J_2})$ contributes to $CP$ violation, and $D$ and $F$ reduce to
\begin{eqnarray}
D&=&-2\sin\phi\Big\{\Big[P_{J_1}P_{J_2}\mathcal{T}_{J_1}\mathcal{P}_{J_2}\sin(\delta_{J_2}-\delta_{J_1})\Big]
+[J_1\leftrightarrow J_2]\Big\}%
,%
\\ 
F&=&\Big\{P_{J_1}^2(\mathcal{T}_{J_1}^2+\mathcal{P}_{J_1}^2)
+P_{J_1}P_{J_2}\Big[\mathcal{T}_{J_1}\mathcal{T}_{J_2}
+\mathcal{P}_{J_1}\mathcal{P}_{J_2}+2\mathcal{T}_{J_1}\mathcal{P}_{J_2}\cos\phi\Big]\cos(\delta_{J_1}-\delta_{J_2})
\Big\}
\nonumber\\&&
+\{J_1\leftrightarrow J_2\}%
.
\end{eqnarray}

In the following of this section, we will focus on the situation when $J_1=0$ and $J_2=1$.
Since the zero point for $P_1\left(g_{s_{12}}(s_{13})\right)$ lies at $s_{13}=\hat{s}_{13}$, this allows us to divide the allowed region of $s_{13}$ into two parts: $\Omega$ and $\bar{\Omega}$, where in $\Omega$ $s_{13}>\hat{s}_{13}$ and in $\bar{\Omega}$ $s_{13}<\hat{s}_{13}$.
The $CP$ asymmetries in the regions $\Omega$ and $\bar{\Omega}$, after integration over $s_{13}$, are found to be
\begin{eqnarray}
A_{CP}^{\Omega}=\frac{\hat{\mathcal{S}}_-^\Omega+\hat{\mathcal{A}}_-^\Omega}{\hat{\mathcal{S}}_+^\Omega+\hat{\mathcal{A}}_+^\Omega},
~~~~~
A_{CP}^{\bar{\Omega}}=\frac{\hat{\mathcal{S}}_-^{\bar{\Omega}}+\hat{\mathcal{A}}_-^{\bar{\Omega}}}{\hat{\mathcal{S}}_+^{\bar{\Omega}}+\hat{\mathcal{A}}_+^{\bar{\Omega}}},
\end{eqnarray}
where
\begin{eqnarray}
\hat{\mathcal{S}}_-^\Omega
&=&
-2\sin\phi\left[\mathcal{T}_0\mathcal{P}_0\sin\alpha_0+\frac{1}{3}\mathcal{T}_1\mathcal{P}_1\sin\alpha_1\right],\\
\hat{\mathcal{S}}_+^\Omega
&=&
\bigg[\mathcal{T}_0^2+\mathcal{P}_0^2+2\mathcal{T}_0\mathcal{P}_0\cos\alpha_0\cos\phi 
+\frac{1}{3}\left(\mathcal{T}_1^2+\mathcal{P}_1^2+2\mathcal{T}_1\mathcal{P}_1\cos\alpha_1\cos\phi\right)\bigg],\\
\hat{\mathcal{A}}_-^\Omega
&=&
\sin\phi\Big[\mathcal{T}_0\mathcal{P}_1\sin(\alpha_1+\delta_1-\delta_0)
+\mathcal{T}_1\mathcal{P}_0\sin(\alpha_0+\delta_0-\delta_1)\Big],\\
\hat{\mathcal{A}}_+^\Omega
&=&
-\Big\{\mathcal{T}_0\mathcal{T}_1\cos(\delta_0-\delta_1)+\mathcal{P}_0\mathcal{P}_1\cos(\alpha_0 \!-\alpha_1\!+\delta_0 \!-\delta_1)\nonumber\\
&&
+\cos\phi\Big[\mathcal{T}_0\mathcal{P}_1\cos(\alpha_1+\delta_1-\delta_0)
+\mathcal{T}_1\mathcal{P}_0\cos(\alpha_0+\delta_0-\delta_1)\Big]\Big\}.
\end{eqnarray}
From the above expressions, one can check that under the interchange of $\Omega$ and $\bar{\Omega}$, $\hat{\mathcal{S}}_\pm^{\bar{\Omega}}$ are symmetric while $\hat{\mathcal{A}}_\pm^{\bar{\Omega}}$ are antisymmetric, i.e.,
\begin{eqnarray}
\hat{\mathcal{S}}_\pm^{\bar{\Omega}}=\hat{\mathcal{S}}_\pm^\Omega,
~~~~
\hat{\mathcal{A}}_\pm^{\bar{\Omega}}=-\hat{\mathcal{A}}_\pm^\Omega.
\end{eqnarray}
Because of the presence of the antisymmetric terms, $CP$ asymmetries in the two regions can be very different.

In practice, the two regions $\Omega$ and $\bar{\Omega}$ are not defined for fixed $s_{12}$.
 In stead, $s_{12}$ lies in a small interval where both of the two resonances are dominant, for example, $\bar{s}_{12}-\lambda_1<s_{12}<\bar{s}_{12}+\lambda_2$ ($\lambda_1$ and $\lambda_2$ are small).
Then the localized $CP$ asymmetry in the region $\omega$ ($\omega=\Omega,\bar{\Omega}$) takes the form
\begin{eqnarray}
A_{CP}^{\omega}=\frac{\int_{\bar{s}_{12}-\lambda_1}^{\bar{s}_{12}+\lambda_2}\text{d}s_{12} (\hat{\mathcal{S}}_-^\omega+\hat{\mathcal{A}}_-^\omega)}{\int_{\bar{s}_{12}-\lambda_1}^{\bar{s}_{12}+\lambda_2}\text{d}s_{12}(\hat{\mathcal{S}}_+^\omega+\hat{\mathcal{A}}_+^\omega)},
\end{eqnarray}
which is exactly the case in Ref. \cite{ZhangGuoYang}.

Besides the $CP$ asymmetry, other quantities may also have interesting behaviors. 
For example, one can check that the quality $R_+$, which is defined as
\begin{equation}
R_+=\frac{\int_\Omega ds_{13}(|\mathcal{M}|^2+|\overline{\mathcal{M}}|^2)-\int_{\bar{\Omega}} ds_{13}(|\mathcal{M}|^2+|\overline{\mathcal{M}}|^2)}{\int_\Omega ds_{13}(|\mathcal{M}|^2+|\overline{\mathcal{M}}|^2)+\int_{\bar{\Omega}} ds_{13}(|\mathcal{M}|^2+|\overline{\mathcal{M}}|^2)}
\end{equation}
equals to $A^\Omega_+/S^\Omega_+$ and is nonzero. Even if the decay process $B\to M_1 M_2 M_3$ is not a weak one, the interference of spin-0 and spin-1 resonances also leads to interesting phenomenology. The quantity $R_+$ is again nonzero. 
The nonzero value of $R_+$ is originated from the interference of the spin-0 and spin-1 resonances.
If one (no matter which one) of the resonances dominates over the other one, $R_+$ will equal to zero.

%
\section{\label{sec:Application}Application to $B^\pm\to K^\pm\pi^+\pi^-$}
%
In this section, we will apply the mechanism which was considered in  last section to the decay  $B^\pm\to K^\pm \pi^+\pi^-$.
We will show that  the interference of the two resonances, $f_0(500)$ and $\rho^0(770)$, which are spin-0 and spin-1, respectively, can lead to large localized $CP$ asymmetry difference around the vicinity of $f_0(500)$ in the phase space.
The corresponding effective Hamiltonian can be expressed as  \cite{Buchalla:1995vs}
\begin{equation}
\mathcal{H}_{\text{eff}} =
\frac {G_{F}}{\sqrt 2} \biggl[  V_{ub}V_{uq}^{\ast}\bigl( C_{1} O_{1}^{u} + 
C_{2}O_{2}^{u} \bigr)+V_{cb}V_{cq}^\ast\bigl( C_{1} O_{1}^{c} + 
C_{2}O_{2}^{c} \bigr)
-  V_{tb}V_{tq}^{\ast} \sum_{i=3}^{10} C_{i}O_{i} \biggr]+ h.c.\ ,
\end{equation}
where $G_{F}$ is the Fermi constant, $V_{qq'}$ is the CKM matrix element, $C_{i}(\mu)$ ($i=1,\cdots,10$)  are the Wilson coefficients, $O_i(\mu)$ are the operators from Operator Product Expansion, $\mu$ is the typical energy scale for the decay process. 
The local four quark operators $O_{i}$ can be written as
\begin{align}
O_{1}^{q'}& = \bar{q}_{\alpha} \gamma_{\mu}(1-\gamma{_5})q'_{\beta}\bar{q}'_{\beta} \gamma^{\mu}(1-\gamma{_5})
b_{\alpha}\ , & O_{2}^{q'}& = \bar{q} \gamma_{\mu}(1-\gamma{_5})q'\bar{q}' \gamma^{\mu}(1-\gamma{_5})b\ ,  \nonumber \\
O_{3}& = \bar{q} \gamma_{\mu}(1-\gamma{_5})b \sum_{q\prime}\bar{q}^{\prime}\gamma^{\mu}(1-\gamma{_5})
q^{\prime}\ , & O_{4}& =\bar{q}_{\alpha} \gamma_{\mu}(1-\gamma{_5})b_{\beta} 
\sum_{q\prime}\bar{q}^{\prime}_{\beta}\gamma^{\mu}(1-\gamma{_5})q^{\prime}_{\alpha}\ , \nonumber \\
O_{5}& =\bar{q} \gamma_{\mu}(1-\gamma{_5})b \sum_{q'}\bar{q}^
{\prime}\gamma^{\mu}(1+\gamma{_5})q^{\prime}\ , & O_{6}& =\bar{q}_{\alpha} \gamma_{\mu}(1-\gamma{_5})b_{\beta} 
\sum_{q'}\bar{q}^{\prime}_{\beta}\gamma^{\mu}(1+\gamma{_5})q^{\prime}_{\alpha}\ ,  \nonumber \\ 
O_{7}& =\frac{3}{2}\bar{q} \gamma_{\mu}(1-\gamma{_5})b \sum_{q'}e_{q^{\prime}}
\bar{q}^{\prime} \gamma^{\mu}(1+\gamma{_5})q^{\prime}\ , & O_{8}& =\frac{3}{2}\bar{q}_{\alpha} 
\gamma_{\mu}(1-\gamma{_5})b_{\beta} 
\sum_{q'}e_{q^{\prime}}\bar{q}^{\prime}_{\beta}\gamma^{\mu}(1+\gamma{_5})q^{\prime}_{\alpha}\ , \nonumber \\
O_{9}& =\frac{3}{2}\bar{q} \gamma_{\mu}(1-\gamma{_5})b \sum_{q'}e_{q^{\prime}}
\bar{q}^{\prime} \gamma^{\mu}(1-\gamma{_5})q^{\prime}\ , & O_{10}& =\frac{3}{2}\bar{q}_{\alpha}
 \gamma_{\mu}(1-\gamma{_5})b_{\beta} \sum_{q'}e_{q^{\prime}}\bar{q}^{\prime}_{\beta}
\gamma^{\mu}(1-\gamma{_5})q^{\prime}_{\alpha}\ ,\nonumber
\end{align}
where $\alpha$ and $\beta$ represent color indices, $e_{q'}$ is the charge of the quark $q'$ in unit of  the absolute electron charge.
With the effective Hamiltonian at hand, we can derive the matrix element for $B^-\to\rho^0\pi^-$ and $B^-\to f_0(500)\pi^-$ via the factorization approach.

We also need the effective Hamiltonian for $\rho^0\to\pi^+\pi^-$ and $f_0(500)\to\pi^+\pi^-$, which can be formally expressed as 
\begin{eqnarray}
\mathcal{H}_{\rho^0\pi\pi}&=&-ig_{\rho\pi\pi}\rho^0_{\mu}\pi^+\overleftrightarrow{\partial} \pi^-,\\
\mathcal{H}_{f_0\pi\pi}&=&g_{f_0\pi\pi}f_0(2\pi^+\pi^-+\pi^0\pi^0),
\end{eqnarray}
where $\rho^0_\mu$, $f_0$ and $\pi^\pm$ are the field operators for $\rho^0$, $f_0(500)$ and $\pi$ mesons, $g_{\rho\pi\pi}$ and $g_{f_0\pi\pi}$ are the effective coupling constants, which should be in principle determined by the underling theory, i.e., QCD. 
The effective coupling constants can be expressed in terms of the decay constants:
\begin{eqnarray}
g_{\rho\pi\pi}^2&=&\frac{48\pi}{\Big(1-\frac{4m_\pi^2}{m_\rho^2}\Big)^{3/2}}\cdot\frac{\Gamma_{\rho^0\to\pi^+\pi^-}}{m_\rho},\\
g_{f_0\pi\pi}^2&=&\frac{4\pi m_{f_0}\Gamma_{f_0\to\pi^+\pi^-}}{\Big(1-\frac{4m_\pi^2}{m_{f_0}^2}\Big)^{1/2}}.
\end{eqnarray}
Both $f_0(500)$ and $\rho^0(770)$ decay into one pion pair dominantly. 
One can easily check that $\Gamma_{\rho^0}\simeq\Gamma_{\rho^0\to\pi^+\pi^-}$, and $\Gamma_{f_0}\simeq\frac{3}{2}\Gamma_{f_0\to\pi^+\pi^-}$

The vector meson $\rho^0(770)$ are usually the dominant resonance for $B$ meson decay channels  with one $\pi^+\pi^-$ pair in the final state, while $f_0(500)$ is not.
This makes both the two resonances, $f_0(500)$ and $\rho^0(770)$, are dominant when the invariant mass of the $\pi^+\pi^-$ pair is around the mass of $f_0(500)$.
As a result, the decay amplitude for $B^- \to K^- \pi^+\pi^-$ can be expressed as
\begin{equation}
\mathcal{M}_{B^- \to K^- \pi^+\pi^-}=\mathcal{M}_{f_0}+\mathcal{M}_{\rho^0} e^{i\tilde{\delta}},
\end{equation}
when the invariant mass of the $\pi^+\pi^-$ pair is around the vicinity of $f_0(500)$, where $\mathcal{M}_{f_0(\rho^0)}$ is the transition amplitude for the cascade decay $B^-\to K^- f_0(\rho^0)$, $f_0(\rho^0)\to \pi^+\pi^-$, $\tilde{\delta}$ is the relative strong phase between $\mathcal{M}_{\rho^0}$ and $\mathcal{M}_{f_0}$ .

With the effective Hamiltonians at hand, one can in principle calculate the transition amplitude via   the QCD factorization approach \cite{Beneke:1999br} or perturbative QCD approach \cite{Li:1994cka}, etc..
These approaches will generate complex phases in the effective Wilson coefficients.
However, these strong phases usually result in a small net strong phase between the penguin  and tree parts of the amplitude.  
Besides, since we are working in the vicinity of $f_0(500)$,  any factorization approach seems not to be accurate for $B^\pm\to K^\pm\rho^0$ when $\rho^0$ is off shell.
In view of this, we will use a naive factorization approach for both  $B^\pm\to K^\pm\rho^0$ and  $B^\pm\to K^\pm f_0(500)$.
As a result, the amplitudes take the form
\footnote{Just as the case of $B^\pm\to\pi^\pm\pi^+\pi^-$ in Ref. \cite{ZhangGuoYang}, for the decay $B^\pm\to K^\pm\pi^+\pi^-$, there are also annihilation terms which are also chiral enhancement terms in the meantime. However, these terms are about four times smaller for $B^\pm\to K^\pm\pi^+\pi^-$ than $B^\pm\to \pi^\pm\pi^+\pi^-$. Because of this, we simply neglect these terms here.}
\begin{widetext}
\begin{eqnarray}
\mathcal{M}_{\rho^0}&=&\frac{2m_\rho g_{\rho\pi\pi}(\hat{s}_{K^-\pi^+}-s_{K^-\pi^+})}{s-m_{\rho^2}+im_\rho\Gamma_\rho}\cdot\Big\{ V_{ub}V_{us}^\ast\Big[\frac{1}{\sqrt{2}}a_1f_\rho F_1+a_2f_KA_0\Big]
\nonumber\\&&
-V_{tb}V_{ts}^\ast\Big[\frac{3}{2\sqrt{2}}(a_7+a_9)f_\rho F_1+\Big(a_4+a_{10}-\frac{2(a_6+a_8)m_K^2}{(m_s+m_u)(m_b+m_u)}\Big)f_KA_0\Big]\Big\},\\
\mathcal{M}_{f_0}&=&\frac{2g_{f_0\pi\pi}}{s-m_{f_0^2}+im_{f_0}\Gamma_{f_0}}f_\pi m_B^2F_0^{B\to f_0}(m_K^2)
\nonumber\\&&
\cdot\Big\{ V_{ub}V_{us}^\ast a_2
-V_{tb}V_{ts}^\ast\Big[(a_4+a_{10})-\frac{2(a_6+a_8)m_K^2}{(m_s+m_u)(m_b+m_u)}\Big]\Big\},
\end{eqnarray}
\end{widetext}
where $F_1$ and $A_0$ are short for the form factors $F_1^{(B\to K)}(m_\rho^2)$ and $A_0^{(B\to \rho)}(m_K^2)$, respectively, all the $a_i$'s are built up from the Wilson coefficients $C_{i}$'s, and take the form $a_i=C_i+C_{i+1}/N_c$ for odd $i$ and $a_i=C_i+C_{i-1}/N_c$ for even $i$.
In deriving the above expression for the amplitudes, we have assumed that both $f_{0}(500)$ and $\rho^0(770)$ do not have  the $s\bar{s}$ component (or at least negligible).
This is a rough estimation, especially for $f_0(500)$,
because the structure of $f_0(500)$ is still unclear
\footnote{ 
Theoretical analysis shows that it has a large $qq\bar{q}\bar{q}$ component \cite{Schechter:2012zc}.}.

We use a set of Wilson coefficients from Ref. \cite{Buchalla:1995vs}:
\begin{eqnarray}
&&C_1=-0.185, ~C_2=1.082, ~C_3=0.014, ~C_4=-0.035, 
\nonumber\\
&&
C_5=0.009, ~C_6=-0.041, ~C_7=-0.002\alpha, 
\nonumber\\
&&
C_8=0.054\alpha, ~C_9=-1.292\alpha, ~C_{10}=0.263\alpha,\nonumber
\end{eqnarray}
where $\alpha$ is the fine structure constant and all the Wilson coefficients are taken in the naive dimensional regularization scheme for $\mu=\overline{m}_b(m_b)=4.40$ GeV, $m_t=170$ GeV, $\Lambda_{\overline{MS}}^{(5)}=225$ MeV.
We also need three form factors, $F_1^{(B^-\to K^-)}$, $A_0^{(B^-\to\rho^0)}$ and $F_0^{(B^-\to f_0)}$.
In our numerical calculation, we use \cite{Cheng:2003sm}
\begin{eqnarray}
F_1^{(B\to K)}(0)=0.35,\\
A_0^{(B\to\rho)}(0)=0.28.
\end{eqnarray}
Since most of the models indicate that the $B$ meson to a light meson form factor at zero recoil lies around 0.3, we simply set
\begin{equation}
F_0^{(B\to f_0)}(0)=0.3.
\end{equation}
One of the commonly used approximations for these form factors is the monopole approximation:
\begin{equation}
f(s)=\frac{f(0)}{1-\frac{s}{m_{\text{p}}^2}},
\end{equation}
where $f=F_1^{(B\to K)}$, $A_0^{(B\to\rho)}$, or $F_0^{(B\to f_0)}$, $m_\text{p}$ is the pole mass.
The pole mass should be different for different form factors (around 5 to 6 GeV).
However, since $s_L$ and $m_\pi^2$ are small compared with the pole mass squared, we will simply replace $f(s)$ or $f(m_K^2)$ by $f(0)$.

We confront with two resonances, $\rho^0(770)$ and $f_0(500)$. 
The masses and total decay widths of these two resonances in our numerical calculation are (in GeV) \cite{Beringer:1900zz}
\begin{eqnarray}
&&m_{\rho^0(770)}=0.775, ~~~~\Gamma_{\rho^0(770)}=0.149,\nonumber\\
&&m_{f_0(500)}=0.500, ~~~~\Gamma_{f_0(500)}=0.500.\nonumber
\end{eqnarray}

\begin{figure}[htbp]
\begin{center}
\includegraphics[width=0.8\textwidth]{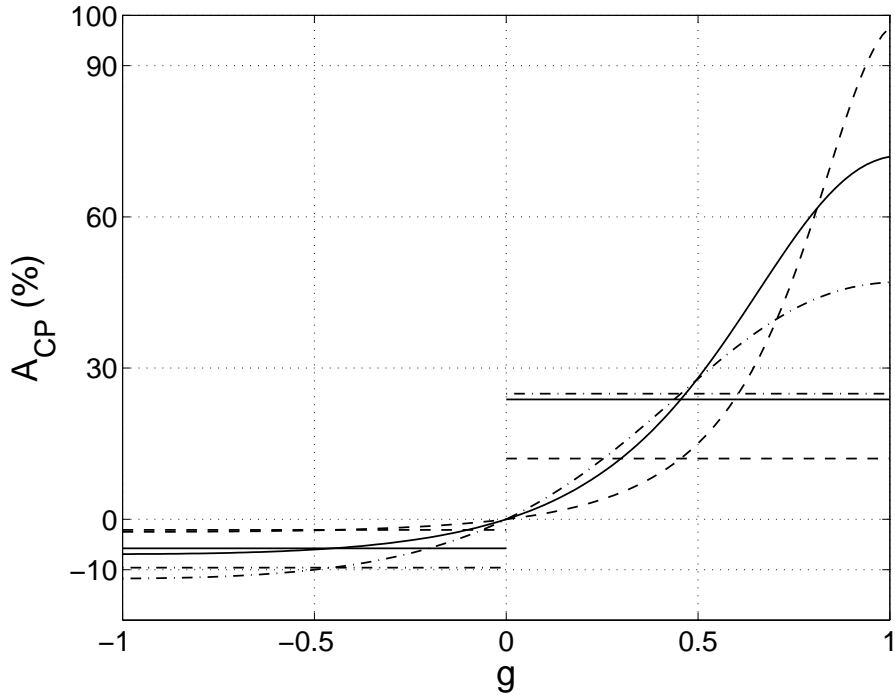}
\caption{The differential $CP$ asymmetry (curved lines) as a function of $g=g_{m_{f_0}^2}(s_{K^\mp\pi^\pm})$. We also show the localized $CP$ asymmetries (straight lines) averaged over the regions $\Omega$ and $\bar{\Omega}$, respectively. Dash-dotted lines, solid lines, and dashed lines are for $\tilde{\delta}=200^\circ$, $220^\circ$ and $240^\circ$, respectively.
}
\label{Fig:ACP}
\end{center}
\end{figure}

With all the above considerations,  one can see that we have only one free parameter, which is the strong phase $\tilde{\delta}$.
The latest experimental $CP$ asymmetry data for the decay channel $B^\pm\to K^\pm \pi^+\pi^-$ is from LHCb Collaboration \cite{LHCb-CONF-2012-018}. 
Their experimental results indicate that when the invariant mass of $\pi^+\pi^-$ is around the vicinity of $f_0(500)$, the $CP$ asymmetry can be larger than about 30\% for small invariant mass of the $K^\mp\pi^\pm$ pair, and lies between 0 to -10\% for large invariant mass of the $K^\mp\pi^\pm$ pair.
These experimental constraints imply that the strong phase $\tilde{\delta}$ should be between $200^\circ$ and $249^\circ$.
In FIG. \ref{Fig:ACP}, we show the differential CP asymmetry as a function of $g_s(s_{K^\mp\pi^\pm})$ when $s=m_{f_0}^2$ for $\tilde{\delta}=200^\circ$, $220^\circ$, and $240^\circ$, respectively.
One can see that when $g$ is smaller than 0 (corresponding to $\hat{s}_{K^\mp\pi^\pm}<s_{K^\mp\pi^\pm}<s_{K^\mp\pi^\pm,\text{max}}$, which is just the region $\Omega$), the $CP$ asymmetry is very small, while when $g$ is larger than about 0.5 (corresponding to $s_{K^\mp\pi^\pm,\text{min}}<s_{K^\mp\pi^\pm}<\hat{s}_{K^\mp\pi^\pm}-0.5\Delta_{K^\mp\pi^\pm}$), the $CP$ asymmetry becomes very large. 
This is exactly what the LHCb experimental results showed.

%
\section{conclusion\label{sec:Conclusion}}
%
In this paper, we presented a general analysis of the mechanism that induces $CP$ violation by the interference of two resonances with different spins.
We applied this mechanism to the decay process $B^\pm\to K^\pm \pi^+\pi^-$.
When the invariant mass of the $\pi^+\pi^-$ pair is around the vicinity of $f_0(500)$, we found that  a large $CP$ asymmetry difference may exist between large and small invariant masses of the $K^\pm\pi^\mp$ system.
A key observation of this large $CP$ asymmetry difference is that it can be interpreted as the interference of the amplitudes induced by $\rho^0$ and $f_0(500)$ as the intermediate resonances, respectively.

Unlike the $B^\pm\to \pi^\pm\pi^+\pi^-$ case where there are up to five free parameters \cite{ZhangGuoYang}, we have only one parameter for the case $B^\pm\to K^\pm \pi^+\pi^-$, which is the relative strong phase $\tilde{\delta}$.
This makes our analysis of the decay channel $B^\pm\to K^\pm \pi^+\pi^-$ much more simplified.
We found that when the relative strong phase $\tilde{\delta}$ lies between $200^\circ$ and $249^\circ$, theoretical analysis is consistent with the data.

%
\begin{acknowledgments}
This work was partially supported by National Natural Science Foundation of China under Contract No. 10975018, No. 11175020, No. 11225523, No. 11275025 and the Fundamental Research Funds for the Central Universities in China.
One of us (Z.H.Z.) also thank the hospitality of Professor Xin-Nian Wang at Huazhong Normal University.
\end{acknowledgments}

\bibliography{zzhbib/zzh}

\begin{thebibliography}{16}%
\makeatletter
\providecommand \@ifxundefined [1]{%
 \@ifx{#1\undefined}
}%
\providecommand \@ifnum [1]{%
 \ifnum #1\expandafter \@firstoftwo
 \else \expandafter \@secondoftwo
 \fi
}%
\providecommand \@ifx [1]{%
 \ifx #1\expandafter \@firstoftwo
 \else \expandafter \@secondoftwo
 \fi
}%
\providecommand \natexlab [1]{#1}%
\providecommand \enquote  [1]{``#1''}%
\providecommand \bibnamefont  [1]{#1}%
\providecommand \bibfnamefont [1]{#1}%
\providecommand \citenamefont [1]{#1}%
\providecommand \href@noop [0]{\@secondoftwo}%
\providecommand \href [0]{\begingroup \@sanitize@url \@href}%
\providecommand \@href[1]{\@@startlink{#1}\@@href}%
\providecommand \@@href[1]{\endgroup#1\@@endlink}%
\providecommand \@sanitize@url [0]{\catcode `\\12\catcode `\$12\catcode
  `\&12\catcode `\#12\catcode `\^12\catcode `\_12\catcode `\%12\relax}%
\providecommand \@@startlink[1]{}%
\providecommand \@@endlink[0]{}%
\providecommand \url  [0]{\begingroup\@sanitize@url \@url }%
\providecommand \@url [1]{\endgroup\@href {#1}{\urlprefix }}%
\providecommand \urlprefix  [0]{URL }%
\providecommand \Eprint [0]{\href }%
\providecommand \doibase [0]{http://dx.doi.org/}%
\providecommand \selectlanguage [0]{\@gobble}%
\providecommand \bibinfo  [0]{\@secondoftwo}%
\providecommand \bibfield  [0]{\@secondoftwo}%
\providecommand \translation [1]{[#1]}%
\providecommand \BibitemOpen [0]{}%
\providecommand \bibitemStop [0]{}%
\providecommand \bibitemNoStop [0]{.\EOS\space}%
\providecommand \EOS [0]{\spacefactor3000\relax}%
\providecommand \BibitemShut  [1]{\csname bibitem#1\endcsname}%
\let\auto@bib@innerbib\@empty
\bibitem [{\citenamefont {Christenson}\ \emph {et~al.}(1964)\citenamefont
  {Christenson}, \citenamefont {Cronin}, \citenamefont {Fitch},\ and\
  \citenamefont {Turlay}}]{Christenson:1964fg}%
  \BibitemOpen
  \bibfield  {author} {\bibinfo {author} {\bibfnamefont {J.}~\bibnamefont
  {Christenson}}, \bibinfo {author} {\bibfnamefont {J.}~\bibnamefont {Cronin}},
  \bibinfo {author} {\bibfnamefont {V.}~\bibnamefont {Fitch}}, \ and\ \bibinfo
  {author} {\bibfnamefont {R.}~\bibnamefont {Turlay}},\ }\href {\doibase
  10.1103/PhysRevLett.13.138} {\bibfield  {journal} {\bibinfo  {journal} {Phys.
  Rev. Lett.}\ }\textbf {\bibinfo {volume} {13}},\ \bibinfo {pages} {138}
  (\bibinfo {year} {1964})}\BibitemShut {NoStop}%
\bibitem [{\citenamefont {Cabibbo}(1963)}]{Cabibbo:1963yz}%
  \BibitemOpen
  \bibfield  {author} {\bibinfo {author} {\bibfnamefont {N.}~\bibnamefont
  {Cabibbo}},\ }\href {\doibase 10.1103/PhysRevLett.10.531} {\bibfield
  {journal} {\bibinfo  {journal} {Phys. Rev. Lett.}\ }\textbf {\bibinfo
  {volume} {10}},\ \bibinfo {pages} {531} (\bibinfo {year} {1963})}\BibitemShut
  {NoStop}%
\bibitem [{\citenamefont {Kobayashi}\ and\ \citenamefont
  {Maskawa}(1973)}]{Kobayashi:1973fv}%
  \BibitemOpen
  \bibfield  {author} {\bibinfo {author} {\bibfnamefont {M.}~\bibnamefont
  {Kobayashi}}\ and\ \bibinfo {author} {\bibfnamefont {T.}~\bibnamefont
  {Maskawa}},\ }\href {\doibase 10.1143/PTP.49.652} {\bibfield  {journal}
  {\bibinfo  {journal} {Prog. Theor. Phys.}\ }\textbf {\bibinfo {volume}
  {49}},\ \bibinfo {pages} {652} (\bibinfo {year} {1973})}\BibitemShut
  {NoStop}%
\bibitem [{\citenamefont {Carter}\ and\ \citenamefont
  {Sanda}(1981)}]{Carter:1980tk}%
  \BibitemOpen
  \bibfield  {author} {\bibinfo {author} {\bibfnamefont {A.~B.}\ \bibnamefont
  {Carter}}\ and\ \bibinfo {author} {\bibfnamefont {A.}~\bibnamefont {Sanda}},\
  }\href {\doibase 10.1103/PhysRevD.23.1567} {\bibfield  {journal} {\bibinfo
  {journal} {Phys. Rev.}\ }\textbf {\bibinfo {volume} {D23}},\ \bibinfo {pages}
  {1567} (\bibinfo {year} {1981})}\BibitemShut {NoStop}%
\bibitem [{\citenamefont {Bigi}\ and\ \citenamefont
  {Sanda}(1981)}]{Bigi:1981qs}%
  \BibitemOpen
  \bibfield  {author} {\bibinfo {author} {\bibfnamefont {I.~I.}\ \bibnamefont
  {Bigi}}\ and\ \bibinfo {author} {\bibfnamefont {A.}~\bibnamefont {Sanda}},\
  }\href {\doibase 10.1016/0550-3213(81)90519-8} {\bibfield  {journal}
  {\bibinfo  {journal} {Nucl. Phys.}\ }\textbf {\bibinfo {volume} {B193}},\
  \bibinfo {pages} {85} (\bibinfo {year} {1981})}\BibitemShut {NoStop}%
\bibitem [{\citenamefont {Aaij}\ \emph
  {et~al.}(2012{\natexlab{a}})\citenamefont {Aaij} \emph
  {et~al.}}]{LHCb-CONF-2012-018}%
  \BibitemOpen
  \bibfield  {author} {\bibinfo {author} {\bibfnamefont {R.}~\bibnamefont
  {Aaij}} \emph {et~al.} (\bibinfo {collaboration} {{LHCb Collaboration}}),\
  }\href@noop {} {} (\bibinfo {year} {2012}{\natexlab{a}}),\ \bibinfo {note}
  {{LHCb-CONF-2012-018}}\BibitemShut {NoStop}%
\bibitem [{\citenamefont {Aaij}\ \emph
  {et~al.}(2012{\natexlab{b}})\citenamefont {Aaij} \emph
  {et~al.}}]{LHCb-CONF-2012-028}%
  \BibitemOpen
  \bibfield  {author} {\bibinfo {author} {\bibfnamefont {R.}~\bibnamefont
  {Aaij}} \emph {et~al.} (\bibinfo {collaboration} {{LHCb Collaboration}}),\
  }\href@noop {} {} (\bibinfo {year} {2012}{\natexlab{b}}),\ \bibinfo {note}
  {{LHCb-CONF-2012-028}}\BibitemShut {NoStop}%
\bibitem [{\citenamefont {de~Miranda}(2013)}]{deMiranda:2013kg}%
  \BibitemOpen
  \bibfield  {author} {\bibinfo {author} {\bibfnamefont {J.~M.}\ \bibnamefont
  {de~Miranda}} (\bibinfo {collaboration} {LHCb Collaboration}),\ }\href@noop
  {} {\  (\bibinfo {year} {2013})},\ \Eprint {http://arxiv.org/abs/1301.0283}
  {arXiv:1301.0283 [hep-ex]} \BibitemShut {NoStop}%
\bibitem [{\citenamefont {Zhang}\ \emph {et~al.}(2013)\citenamefont {Zhang},
  \citenamefont {Guo},\ and\ \citenamefont {Yang}}]{ZhangGuoYang}%
  \BibitemOpen
  \bibfield  {author} {\bibinfo {author} {\bibfnamefont {Z.-H.}\ \bibnamefont
  {Zhang}}, \bibinfo {author} {\bibfnamefont {X.-H.}\ \bibnamefont {Guo}}, \
  and\ \bibinfo {author} {\bibfnamefont {Y.-D.}\ \bibnamefont {Yang}},\ }\href
  {\doibase 10.1103/PhysRevD.87.076007} {\bibfield  {journal} {\bibinfo
  {journal} {Phys. Rev.}\ }\textbf {\bibinfo {volume} {D87}},\ \bibinfo {pages}
  {076007} (\bibinfo {year} {2013})},\ \Eprint {http://arxiv.org/abs/1303.3676}
  {arXiv:1303.3676 [hep-ph]} \BibitemShut {NoStop}%
\bibitem [{\citenamefont {Chung}()}]{Chung:1971ri}%
  \BibitemOpen
  \bibfield  {author} {\bibinfo {author} {\bibfnamefont {S.~U.}\ \bibnamefont
  {Chung}},\ }\href@noop {} {\enquote {\bibinfo {title} {Spin formalisms},}\
  }\bibinfo {note} {{CERN Yellow Report: CERN-71-08 (1971)}}\BibitemShut
  {NoStop}%
\bibitem [{\citenamefont {Buchalla}\ \emph {et~al.}(1996)\citenamefont
  {Buchalla}, \citenamefont {Buras},\ and\ \citenamefont
  {Lautenbacher}}]{Buchalla:1995vs}%
  \BibitemOpen
  \bibfield  {author} {\bibinfo {author} {\bibfnamefont {G.}~\bibnamefont
  {Buchalla}}, \bibinfo {author} {\bibfnamefont {A.~J.}\ \bibnamefont {Buras}},
  \ and\ \bibinfo {author} {\bibfnamefont {M.~E.}\ \bibnamefont
  {Lautenbacher}},\ }\href {\doibase 10.1103/RevModPhys.68.1125} {\bibfield
  {journal} {\bibinfo  {journal} {Rev. Mod. Phys.}\ }\textbf {\bibinfo {volume}
  {68}},\ \bibinfo {pages} {1125} (\bibinfo {year} {1996})},\ \Eprint
  {http://arxiv.org/abs/hep-ph/9512380} {arXiv:hep-ph/9512380} \BibitemShut
  {NoStop}%
\bibitem [{\citenamefont {Beneke}\ \emph {et~al.}(1999)\citenamefont {Beneke},
  \citenamefont {Buchalla}, \citenamefont {Neubert},\ and\ \citenamefont
  {Sachrajda}}]{Beneke:1999br}%
  \BibitemOpen
  \bibfield  {author} {\bibinfo {author} {\bibfnamefont {M.}~\bibnamefont
  {Beneke}}, \bibinfo {author} {\bibfnamefont {G.}~\bibnamefont {Buchalla}},
  \bibinfo {author} {\bibfnamefont {M.}~\bibnamefont {Neubert}}, \ and\
  \bibinfo {author} {\bibfnamefont {C.~T.}\ \bibnamefont {Sachrajda}},\ }\href
  {\doibase 10.1103/PhysRevLett.83.1914} {\bibfield  {journal} {\bibinfo
  {journal} {Phys. Rev. Lett.}\ }\textbf {\bibinfo {volume} {83}},\ \bibinfo
  {pages} {1914} (\bibinfo {year} {1999})},\ \Eprint
  {http://arxiv.org/abs/hep-ph/9905312} {arXiv:hep-ph/9905312} \BibitemShut
  {NoStop}%
\bibitem [{\citenamefont {Li}\ and\ \citenamefont {Yu}(1995)}]{Li:1994cka}%
  \BibitemOpen
  \bibfield  {author} {\bibinfo {author} {\bibfnamefont {H.-n.}\ \bibnamefont
  {Li}}\ and\ \bibinfo {author} {\bibfnamefont {H.-L.}\ \bibnamefont {Yu}},\
  }\href {\doibase 10.1103/PhysRevLett.74.4388} {\bibfield  {journal} {\bibinfo
   {journal} {Phys. Rev. Lett.}\ }\textbf {\bibinfo {volume} {74}},\ \bibinfo
  {pages} {4388} (\bibinfo {year} {1995})},\ \Eprint
  {http://arxiv.org/abs/hep-ph/9409313} {arXiv:hep-ph/9409313} \BibitemShut
  {NoStop}%
\bibitem [{\citenamefont {Schechter}(2012)}]{Schechter:2012zc}%
  \BibitemOpen
  \bibfield  {author} {\bibinfo {author} {\bibfnamefont {J.}~\bibnamefont
  {Schechter}},\ }\href {\doibase 10.1143/PTPS.197.64} {\bibfield  {journal}
  {\bibinfo  {journal} {Prog. Theor. Phys. Suppl.}\ }\textbf {\bibinfo {volume}
  {197}},\ \bibinfo {pages} {64} (\bibinfo {year} {2012})},\ \Eprint
  {http://arxiv.org/abs/1202.3176} {arXiv:1202.3176 [hep-ph]} \BibitemShut
  {NoStop}%
\bibitem [{\citenamefont {Cheng}\ \emph {et~al.}(2004)\citenamefont {Cheng},
  \citenamefont {Chua},\ and\ \citenamefont {Hwang}}]{Cheng:2003sm}%
  \BibitemOpen
  \bibfield  {author} {\bibinfo {author} {\bibfnamefont {H.-Y.}\ \bibnamefont
  {Cheng}}, \bibinfo {author} {\bibfnamefont {C.-K.}\ \bibnamefont {Chua}}, \
  and\ \bibinfo {author} {\bibfnamefont {C.-W.}\ \bibnamefont {Hwang}},\ }\href
  {\doibase 10.1103/PhysRevD.69.074025} {\bibfield  {journal} {\bibinfo
  {journal} {Phys. Rev.}\ }\textbf {\bibinfo {volume} {D 69}},\ \bibinfo
  {pages} {074025} (\bibinfo {year} {2004})},\ \Eprint
  {http://arxiv.org/abs/hep-ph/0310359} {arXiv:hep-ph/0310359} \BibitemShut
  {NoStop}%
\bibitem [{\citenamefont {Beringer}\ \emph {et~al.}(2012)\citenamefont
  {Beringer} \emph {et~al.}}]{Beringer:1900zz}%
  \BibitemOpen
  \bibfield  {author} {\bibinfo {author} {\bibfnamefont {J.}~\bibnamefont
  {Beringer}} \emph {et~al.} (\bibinfo {collaboration} {Particle Data Group}),\
  }\href {\doibase 10.1103/PhysRevD.86.010001} {\bibfield  {journal} {\bibinfo
  {journal} {Phys. Rev.}\ }\textbf {\bibinfo {volume} {D 86}},\ \bibinfo
  {pages} {010001} (\bibinfo {year} {2012})}\BibitemShut {NoStop}%
\end{thebibliography}%

%
\end{document}